\documentstyle[twocolumn,graphicx,aps]{revtex}

\begin{document}

\draft

\title{Creating macroscopic atomic EPR states from Bose condensates}

\author{H. Pu and P. Meystre}
\address{
{Optical Sciences Center, The University of Arizona,
Tucson, AZ 85721}}

\date{\today}
\maketitle
\begin{abstract}

We present a scheme for creating quantum entangled atomic states
through the coherent spin-exchange collision of a spinor
Bose-Einstein condensate. The state generated possesses
macroscopic Einstein-Podolsky-Rosen correlation and the
fluctuation in one of its quasi-spin components vanishes. We show
that an elongated condensate with large aspect ratio is most
suitable for creating such a state.

\end{abstract}

Quantum entanglement lies at the heart of the profound difference
between quantum mechanics and classical physics\cite{bell}. The entanglement
between the states of space-like separated particles is the
fundamental reason for the violation of Bell inequality, and
causes many of the ``paradoxes''  of quantum physics. In recent
years, there has been an interesting maturing of the discussions
of entanglement away from the foundations of quantum mechanics and
to ``applications'' in the emerging field of quantum information
processing.

A majority of the experimental realizations of quantum
entanglement to date involve the creation of entangled photon
pairs. Although ideal as carriers of quantum information, photons
are however normally difficult to store for extended periods of
time, in contrast to massive particles. To overcome this
difficulty, progresses have been made to generate
correlated atom-photon pairs\cite{hag,moore}. Recently, much attention
has also been paid to quantum correlated atomic systems,
particularly nonclassical multi-atom
states\cite{wine,ueda,kuz,cas}, as these systems have important
applications in quantum measurement beyond the ``standard quantum
limits'' as well as in quantum computation.

There have already been several proposals to create entangled
atomic ensembles\cite{polzik,lukin} and one of them has recently
been demonstrated experimentally\cite{measure1}. All of these
schemes rely on mapping the nonclassical properties of
electromagnetic waves, e.g., squeezed light, onto the state of an
atomic system. In this Letter, we show that by taking advantage of
coherent spin-exchange ultracold collisions, one can generate
macroscopic atomic Einstein-Podolsky-Rosen (EPR) states\cite{ein} from a
spinor Bose-Einstein condensate without the need of nonclassical
light fields.

\begin{figure}
\begin{center}
    \includegraphics*[width=0.8\columnwidth,
height=0.55\columnwidth]{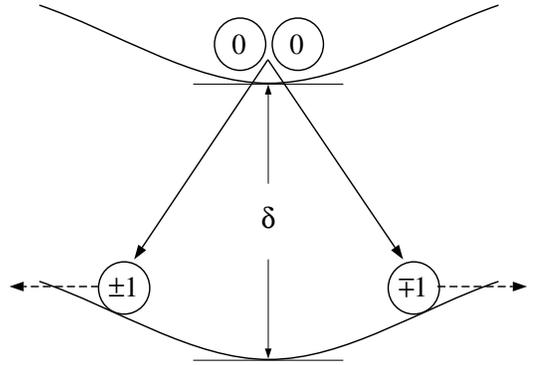}
\vspace{3 mm}
\caption{Entanglement scheme: A spin-0 condensate is
initially prepared. Spin-exchange interaction creates spin-($\pm
1$) atom pairs whose energy level is shifted below that of spin-0
atom by an amount of $\hbar \delta$. This excess energy is
transferred into the kinetic energies of spin-($\pm 1$) atoms
which escape the trap.} \label{fig1}
\end{center}
\end{figure}

We proceed by first giving the general idea of the proposed
technique, and then turn to a more detailed theoretical
discussion. Our scheme is illustrated in Fig.~\ref{fig1}. A spinor
Bose-Einstein condensate consisting of a dilute $F=1$ atomic
sample is initially polarized such that only the spin-0 hyperfine
ground state is populated at time $t=0$. Binary spin-exchange
interaction then convert the spin-0 atoms into pairs of spin-($\pm
1$) atoms. The irreversibility of such a process is provided by
shifting the energy of the spin-0 state above that of the
spin-($\pm 1$) states (which can be achieved using the ac Stark
shift provided by far off-resonant laser light\cite{shift}). As a
result of this detuning, the phase-matching condition, i.e.,
conservation of momentum and energy, ensures that the resultant
atoms in the pair move in opposite directions away from each other
and escape the trap. Quantum entanglement results from our
ignorance about which of the two escaping atoms is in the
spin-(+1) state and which has spin-($-1$).

We now turn to a detailed analysis of this system. At $t=0$, a
condensate of $N_0$ spin-0 atoms is confined in an optical dipole
trap. An additional off-resonant optical field is used to shift
the energy of the spin-0 state above those of the spin-($\pm 1$)
states by an amount $\hbar \delta$ (see Fig.~\ref{fig1}). The
spatial wave function of the condensate, $\varphi({\bf r})$ is
determined by the stationary Gross-Pitaevskii equation.

At $t>0$, the spin-($\pm 1$) states start being populated by the
spin-exchange interaction
\begin{equation}
H=\lambda_a \int \,d{\bf r}\,\hat{\psi}_{+1}^{\dagger}({\bf r},t)
\hat{\psi}_{-1}^{\dagger}({\bf r},t)
\hat{\psi}_0 ({\bf r},t) \hat{\psi}_0({\bf r},t)  + h.c.
\label{h1}
\end{equation}
where $\lambda_a$ is a constant related to the $s$-wave scattering
lengths associated with the hyperfine levels involved, and
$\hat{\psi}_{\alpha}$ is the boson annihilation operator for
spin-$\alpha$ atoms. The effect of atomic recoil during this
process is to transfer the excess energy $\hbar \delta$ into the
kinetic energy of the spin-($\pm 1$) atoms. Therefore, for the
short time scale where the propagation of ($\pm 1$) atoms can be
neglected, we may expand the boson field operators as
\begin{eqnarray}
\hat{\psi}_0({\bf r},t) &=& \varphi({\bf r})\,e^{-i \delta t}
\,\hat{c}_0 (t)\\ \hat{\psi}_{\pm 1,{\bf q}}({\bf r},t) &=&
\varphi({\bf r})\,e^{i({\bf q} \cdot {\bf r}-\omega_q t)}
\,\hat{c}_{\pm 1,{\bf q}}(t),
\end{eqnarray}
where $\omega_q \equiv \hbar |{\bf q}|^2/(2m)$, and the operators
$\{ \hat{c}_{\mu} \}$ obey the boson commutation relations $[
\hat{c}_{\mu},\,\hat{c}_{\nu}^{\dagger} ]=\delta_{\mu,\nu}$. With
these expansions, Hamiltonian (\ref{h1}) may be reexpressed as
\begin{equation}
H=\kappa \int\int\,d{\bf q}\,d{\bf q'}\,\rho({\bf q},{\bf q'})
e^{i\Delta_{q,q'}t} \,\hat{c}^{\dagger}_{+1,{\bf q}}
\hat{c}^{\dagger}_{-1,{\bf q'}} \hat{c}_0 \hat{c}_0 + h.c.,
\label{h2}
\end{equation}
where $\Delta_{q,q'} \equiv (\omega_q + \omega_{q'} -2\delta)$,
$\kappa \equiv \lambda_a V^2/(2 \pi)^6$ with $V$ being the
quantization volume, and
\begin{equation}
\rho({\bf q},{\bf q'}) = \int\,d{\bf r}\,|\varphi ({\bf
r})|^4\,e^{-i({\bf q}+{\bf q'}) \cdot {\bf r}}. \label{rho}
\end{equation}
Eq.~(\ref{h2}) is reminiscent of the Hamiltonian describing
parametric down conversion processes in nonlinear and quantum
optics. As is well known, these processes lead to squeezing and to
the generation of entangled photon pairs.

For short enough interaction times, the population of the
sidemodes ($\pm 1$) remain small compared to $N_0$. In this
regime, we neglect the depletion of the spin-0 state and treat
$\hat{c}_0$ as a $c$-number $c_0$ such that $|c_0|^2=N_0$. We can
furthermore neglect those terms in the Hamiltonian (\ref{h2}) that
describe atom-atom interactions involving only the spin-($\pm 1$)
states. Under these assumptions, the Heisenberg dynamics of the
operators $\hat{c}_{\pm 1, {\bf q}}$ resulting from the
interaction Hamiltonian (\ref{h2}) simplifies to
\begin{eqnarray}
\frac{d}{dt} \hat{c}_{+1,{\bf q}} &=&  -i \kappa \int\,d{\bf q'}\,
\rho({\bf q},{\bf q'}) \,e^{i\Delta_{q,q'}t}\,
\hat{c}^{\dagger}_{-1,{\bf q'}}c_0^2 \label{e+} ,\\ \frac{d}{dt}
\hat{c}_{-1,{\bf q}} &=&  -i \kappa \int\,d{\bf q'}\, \rho({\bf
q},{\bf q'}) \,e^{i\Delta_{q,q'}t}\, \hat{c}^{\dagger}_{+1,{\bf
q'}}c_0^2 .\label{e-}
\end{eqnarray}
To solve these equations, we first formally integrate
Eq.~(\ref{e-}) to get
\begin{eqnarray}
\hat{c}_{-1,{\bf q}}(t) &=& \hat{c}_{-1,{\bf q}}(0)-i\kappa
c_0^2\int\, d{\bf q'}\,\rho({\bf q},{\bf q'})\,\int^t_0 \, d\tau\,
e^{i\Delta_{q,q'} \tau} \nonumber\\ &&\;\;\;\;\;\;\times
\hat{c}^{\dagger}_{+1,{\bf q}}(t-\tau) \nonumber \\ & \approx &
\hat{c}_{-1,{\bf q}}(0)-i\kappa c_0^2\int d{\bf q'}\,\rho({\bf
q},{\bf q'})\,\delta (\Delta_{q,q'}) \,\hat{c}^{\dagger}_{+1,{\bf
q}}(t)  \label{ee-}
\end{eqnarray}
where the Markov approximation has been invoked. Inserting Eq.
(\ref{ee-}) into Eq. (\ref{e+}), we obtain
\begin{equation}
\frac{d}{dt} \hat{c}_{+1,{\bf q}} = \frac{N_0^2}{2} \,G_{{\bf q}}\,
\hat{c}_{+1,{\bf q}}+\hat{f}^{\dagger}_{{\bf q}}(t)
\label{eq+}
\end{equation}
where we have defined the gain parameter
\begin{eqnarray}
G_{{\bf q}} &=& 2\pi \kappa^2\int\,d{\bf q'}\, |\rho({\bf q},{\bf
q'})|^2 \,\delta(\Delta_{q,q'}) \label{gq}
\end{eqnarray}
and the noise operator
\begin{eqnarray*}
\hat{f}^{\dagger}_{{\bf q}}(t) &=& -i \kappa c_0^2\int\,d{\bf
q'}\, \rho({\bf q},{\bf q'}) \,e^{i\Delta_{q,q'}t}\,
\hat{c}^{\dagger}_{-1,{\bf q'}}(0),
\end{eqnarray*}
whose correlation functions are given in the Markov approximation
by
\begin{eqnarray*}
\langle \hat{f}^{\dagger}_{{\bf q}}(t) \,\hat{f}_{{\bf q}}(t')
\rangle &=& 0 ,\\ \langle \hat{f}_{{\bf
q}}(t)\,\hat{f}^{\dagger}_{{\bf q'}}(t') \rangle &=& N_0^2 G_{{\bf
q}}\, \delta({\bf q}-{\bf q'})\, \delta(t-t').
\end{eqnarray*}
It is this noise operator that triggers the populating of
spin-(+1) state from quantum fluctuations. In deriving
Eq.~(\ref{eq+}), we have used the approximation $\rho({\bf q},{\bf
q'}) \rho^*({\bf q'},{\bf q''}) \approx |\rho({\bf q},{\bf q'})|^2
\delta({\bf q}-{\bf q''})$ and neglected the principal part
associated with the definition of the $\delta$-function.

Following a similar procedure, we can derive the equation of
motion for $\hat{c}_{-1,{\bf q}}$ as
\begin{equation}
\frac{d}{dt} \hat{c}_{-1,{\bf q}} = \frac{N_0^2}{2} \,G_{{\bf
q}}\, \hat{c}_{-1,{\bf q}}+\hat{g}^{\dagger}_{{\bf q}}(t) ,
\label{eq-}
\end{equation}
where
\begin{equation}
\hat{g}^{\dagger}_{{\bf q}}(t) = -i \kappa \int\,d{\bf q'}\,
\rho({\bf q},{\bf q'}) \,e^{i\Delta_{q,q'}t}
\hat{c}^{\dagger}_{+1,{\bf q'}}(0) \,c_0^2.
\end{equation}

At this level of approximation, which neglects as we recall the
depletion of the spin-0 mode, the Heisenberg equations of motion
(\ref{eq+}) and (\ref{eq-}) are linear. They can readily be
integrated to give
\begin{eqnarray}
\hat{c}_{+1,{\bf q}}(t) &=& {\cal G}_{\bf q}( t)\,
\hat{c}_{+1,{\bf q}}(0)+
\int^t_0\,d\tau\,{\cal G}_{\bf q}( \tau)\,
\hat{f}^{\dagger}_{{\bf q}}(t-\tau) \\
\hat{c}_{-1,{\bf q}}(t) &=& {\cal G}_{\bf q}(t)\,
\hat{c}_{-1,{\bf q}}(0) +
\int^t_0\,d\tau\,{\cal G}_{\bf q}( \tau)\,
\hat{g}^{\dagger}_{{\bf q}}(t-\tau)
\end{eqnarray}
where ${\cal G}_{\bf q}( t) \equiv \exp (N_0^2 G_{{\bf q}}t/2)$.
From these we can calculate the population in modes $\{ \pm 1,{\bf q} \}$:
\begin{eqnarray*}
N_{\pm 1,{\bf q}}=\langle \hat{c}^{\dagger}_{\pm 1,{\bf q}}
\hat{c}_{\pm 1,{\bf q}} \rangle = \exp (N_0^2G_{{\bf q}} t) -1
\end{eqnarray*}
It is also straightforward to calculate the correlation function
\begin{eqnarray}
{\cal C}_{{\bf q},{\bf q'}} &\equiv &
\langle \hat{c}_{-1,{\bf q}}\,
\hat{c}_{+1,{\bf q'}} \rangle \nonumber \\
&=& -i \kappa \,{\cal G}_{\bf q}(t)\, c_0^2\,
\rho_{{\bf q}}({\bf q'})
\frac{{\cal G}_{{\bf q'}}(t)-e^{i \Delta_{q, q'}t}}
{N_0^2G_{{\bf q'}}/2- i \Delta_{q, q'}}
\label{corr}
\end{eqnarray}
The fact that the $\pm 1$ modes are correlated implies that the
two spin states ($\pm 1$) are entangled. It is obviously desirable
that a spin-($-1$) atoms with momentum $\hbar {\bf q}$ be
correlated to a spin-(+1) atom with well defined momentum $\hbar
{\bf q'}$. \

From the definition (\ref{rho}) of $\rho({\bf q},{\bf q'})$, we
conclude that as long as the spatial size of the condensate wave
function is much larger than the reciprocal length $1/|{\bf q}|$
and $1/|{\bf q'}|$, $\rho({\bf q},{\bf q'})$ is approximately
proportional to a delta-function, $\rho({\bf q},{\bf
q'})\longrightarrow \delta({\bf q}+{\bf q'})$. In other words,
under this condition the two correlated atoms resulting from a
spin-changing collision move in opposite directions. Additionally,
the particle momenta $|{\bf q}|$ and $|{\bf q'}|$ have to satisfy
the conservation of energy condition $ \omega_q +\omega_{q'}
-2\delta \approx 0$, a condition that can be met for a large light
shift $\hbar \delta$. We observe that in addition, a large energy
shift $\hbar \delta$ is also required to produce spin-({$\pm 1$)
atoms with sufficiently large kinetic energy to escape the trap.
As already mentioned, this is required to prevents them from
undergoing a collision resulting in a pair of spin-0 atoms.

\begin{figure}
\begin{center}
    \includegraphics*[width=0.8\columnwidth,
height=0.6\columnwidth]{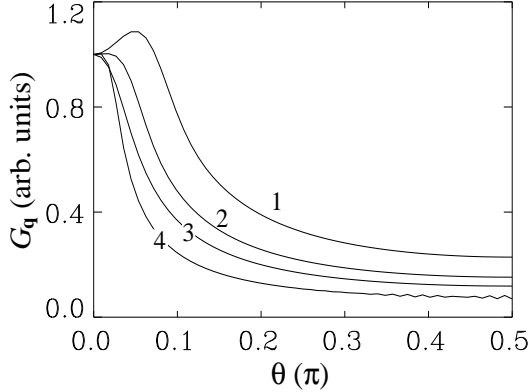}
\vspace{3 mm} \caption{The gain parameter $G_{{\bf q}}$ versus
$\theta$, for a Gaussian and cylindrically symmetric condensate
wave function of the form $\varphi({\bf r}) \propto
\exp[-z^2/(2\sigma_z^2)-(x^2+y^2)/(2\sigma_{\bot}^2)]$. $\theta$
is the angle between ${\bf q}$ and the $z$-axis. In the
calculation, we set $q=|{\bf q}|=\sqrt{2m\delta / \hbar}$. Curve
1: $\sigma_z=10$, $q=10$; curve 2: $\sigma_z=10$, $q=20$; curve 3:
$\sigma_z=10$, $q=40$; curve 4: $\sigma_z=20$, $q=40$. The units
for $\sigma_z$ and $q$ are $\sigma_{\bot}$ and $1/\sigma_{\bot}$,
respectively.} \label{fig2}
\end{center}
\end{figure}

In general, it is not sufficient to just produce entangled atomic
pairs. Rather, one would like to subsequently store them, e.g. in
a dipole trap. It is desirable for this purpose to achieve a high
degree of directionality in the generated atoms, so that they have
a narrow enough angular distribution. To see how this can be
achieved, let us take a closer look at the gain parameter $G_{{\bf
q}}$ appearing in Eq.~(\ref{gq}). The expression of $G_{{\bf q}}$
is reminiscent of a similar gain parameter encountered in the
study of superradiant scattering from a
condensate\cite{supere,supert}. It has been shown in that context
that for a spatially anisotropic condensate, the largest gain
occurs along the longest dimension of the condensate\cite{supert}.
The same conclusion can be reached in the present case.
Fig.~\ref{fig2} illustrates the gain along different directions
for the case of a cylindrically symmetric condensate, for various
light shifts $\hbar \delta$ and aspect ratios. For simplicity, we
choose $q=|{\bf q}|=\sqrt{2m\delta/\hbar}$, and assume that the
condensate has a Gaussian shape. Fig.~2
illustrates quite clearly that a smaller angular distribution of
emitted atoms is obtained for larger aspect ratios and larger $q$.
Thus for a strongly elongated cigar-shaped condensate, the
matter-wave modes along the long axis, which have the largest gain
coefficient $G_{{\bf q}}$, will typically deplete all the
condensate atoms before the population of the off-axis modes can
significantly build up. As a consequence of mode competition, the
emission of the spin-($\pm 1$) atoms is therefore largely confined
to two narrow cones at the two ends of the cigar-shaped
condensate.

From this discussion we conclude that in order to experimentally
realize the proposed scheme,
one should first create an elongated spin-0 condensate with a
large light shift $\hbar \delta$. Spin-exchange interactions then
generate pairs of spin-($\pm 1$) atoms flying in opposite
directions along its long axis. These atoms can be subsequently
captured by two traps located at opposite sides of the original
trap. Eventually, the spin-0 condensate is depleted, with two new
ensembles of pair-wise entangled atoms stored inside the side
traps. We emphasize that although each trap contains both spin-(+1)
and spin-($-1$) atoms, these cannot undergo subsequent spin-exchange
collisions to produce spin-0 atoms, since this process does not
satisfy momentum-energy conservation.

The spin-(${\pm 1}$) atoms being created in pairs, we know for
sure that taken together, the two ensembles must contain equal
number of spin-(+1) and spin-($-1$) atoms --- although how many
spin-(+1) and spin-($-1$) atoms are in each ensemble is unknown.
In the Schr\"{o}dinger picture, such a state may be expressed as
\begin{equation}
|\Psi \rangle=\sum_{m=-N/2}^{N/2}\,a_m\, \left|\frac{N}{2},\,m
\right\rangle_l\, \left| \frac{N}{2},\,-m \right\rangle_r ,
\label{epr}
\end{equation}
where $N/2$ is the total number of atoms in each of the two
``left'' and ``right'' side traps, labeled by $l$ and $r$,
respectively. The integers $m$ and $-m$ represent the difference
in numbers of atoms in the spin states (+1) and ($-1$) in each of the
two traps.

Introducing the $z$-component of the quasi-spin operator
$$\hat{L}_z^{(i)}= \hat{N}^{(i)}_{+1} -\hat{N}^{(i)}_{-1} ,$$
where $\hat{N}_{\pm 1}^{(i)}$ is the number operator for
state-($\pm 1$) in ensemble $i$ and $i=l,r$, we have that
$$
\hat{L}_z^{(i)}\,\left| \frac{N}{2},\,m \right\rangle_i  = m\left|
\frac{N}{2},\,m \right\rangle_i . $$ Since the explicit
expressions of the coefficients $a_m$ in (\ref{epr}) are  unknown,
so are the expectation value and variance for $\hat{L}_z^{(i)}$.
However, a simple calculation shows that
\begin{eqnarray*}
\langle \hat{L}_z \rangle &=& 0 ,\\
( \Delta \hat{L}_z )^2 &=& 0,
\end{eqnarray*}
where $\hat{L}_z \equiv \hat{L}_z^{(l)} + \hat{L}_z^{(r)}$ is the
$z$-component of the {\em total} quasi-spin operator.  Hence,
although the variance of the $\hat{L}_z^{(i)}$ may be large for
the individual ensembles, the variance for the whole system
vanishes. In other words, taken as a whole the two ensembles
represent a maximally spin-squeezed state. This should be
contrasted to the case of $N$ independent atoms in the state
$(|+1\rangle + |-1\rangle)^N$, for which one finds $( \Delta
\hat{L}_z )^2 = N/4$.

We note that if we randomly pick one atom each from the two
side-traps for an atomic ensemble prepared in state (\ref{epr}),
then their degree of entanglement is only of order $1/N$. This is
because although the atoms are created in pairs, we cannot tell
which particular pairs of atoms are entangled. It is only through
the collective spin measurement that the quantum entanglement can
be revealed. The observation of such an macroscopic entanglement
can be carried out with the technique of spectroscopic
detection of collective spin noise at the quantum level described
in Refs.~\cite{measure1,measure2}. In practice, the state
(\ref{epr}) has to be averaged over the statistical distribution
of the total particle number $N$. However, as noted in
Ref.~\cite{polzik}, such fluctuations do not affect the
entanglement significantly for large numbers of atoms.

In conclusion, we have proposed and analyzed a simple scheme to
create a macroscopic EPR-correlated atomic state. 
Such a state possesses a
nonlocal entanglement and is maximally squeezed, in
the sense that the fluctuations of the $z$-component of its
quasi-spin vanish. Hence we believe that this system will have 
important applications in precision measurement as well as in
fundamental physics such as the test of
nonlocality in macroscopic quantum systems. 
Our study shows that an elongated spinor
condensate with large aspect ratio and large energy difference
between spin-0 and spin-($\pm 1$) states is the best candidate to
create such a state. The correlations between the atomic ensembles
arise from the nonlinear atom-atom interaction amongst the
condensate atoms. This distinguishes our work from other proposals
with a similar goal, where the correlations between atoms are
transferred from EPR-correlated light fields. As a consequence,
our scheme can deal with strictly ground state hyperfine atomic
states. This is of considerable advantage, since the entanglement
of the kind described here is therefore robust against decoherence
and immune from the quantum fluctuations caused by the
electromagnetic vacuum field modes, which limit the degree of
entanglement and spin squeezing\cite{polzik,measure1}.

{\em Note}: Upon completion of our work, we noticed a preprint
paper by S{\o}rensen {\em et al.}\cite{zoller} in which 
the possibility of creating squeezed spin state with
Bose condensates is investigated. Our work differs from theirs as
the atomic ensembles we studied are spatially separated with
nonlocal EPR correlation, while theirs does not possess this property.

\acknowledgments
This work is supported in part by the US Office of Naval Research under
Contract No. 14-91-J1205, by the National Science Foundation under Grant
No. PHY98-01099, by the US Army Research Office, and by the Joint Services
Opitcs Program.


\end{document}